# Protecting Human Users Against Cognitive Attacks in Immersive Environments

YAN-MING CHIOU and BOB PRICE, SRI International, USA

CHIEN-CHUNG SHEN and SYED ALI ASIF, University of Delaware, USA

## POSITION STATEMENT

Integrating mixed reality (MR) with artificial intelligence (AI) technologies, including vision, language, audio, reasoning, and planning, enables the AI-powered MR assistant [1] to substantially elevate human efficiency. This enhancement comes from situational awareness, quick access to essential information, and support in learning new skills in the right context throughout everyday tasks. This blend transforms interactions with both the virtual and physical environments, catering to a range of skill levels and personal preferences. For instance, computer vision enables the understanding of the user's environment, allowing for the provision of timely and relevant digital overlays in MR systems. At the same time, language models enhance comprehension of contextual information and support voice-activated dialogue to answer user questions. However, as AI-driven MR systems advance, they also unveil new vulnerabilities, posing a threat to user safety by potentially exposing them to grave dangers [5, 6].

Specifically, MR enhances user interaction with the physical world by overlaying spatial or contextual information onto real-world environments, facilitating a deeper understanding of the surroundings and assigned tasks. This technology helps the user finish the task with contextual background information, spatial information for objects and places, procedural guidance, and supportive widgets designed to enhance concentration and memory retention. Although this advancement augments user experience, MR introduces a spectrum of security issues, as it presents an avenue for malicious entities to manipulate user perceptions through sophisticated visual and auditory deceptions. For instance, utilizing advanced computing capabilities and high-resolution displays inherent in MR devices, malicious entities may leverage computer vision algorithms and Generative AI model [4] to create deceptive overlays. These can hinder correct object recognition, lead users to make incorrect identification and justification, or create the false appearance of objects that don't exist in the user's surroundings. Such manipulations can also alter the perceived attributes of natural objects, potentially leading to operational inaccuracies and user safety.

In addition, MR may incorporate advanced spatial audio features, further immersing users in their digital interactions. The integration of AI-powered audio models [7] capable of producing highly realistic synthetic voices with emotional nuances, achieved with minimal latency and proficient at managing interruptions in conversations, introduces additional attack vectors for exploitation. Strategic placement of spatial sound overlays may mislead, obscure, or implant fictitious auditory perceptions, causing physical harm, psychological distress, or sensory confusion while remaining undetected. The potential for audio-based manipulations to obscure reality, prompt erroneous interpretations, or induce perceptual illusions necessitates a rigorous examination of security protocols within MR environments.

Delving into specifics, certain commercial devices are already equipped with the AI capabilities described above. For instance, Apple Vision Pro can create a unique **3D Persona** [3]. The process involves guiding users with visual cues—using shapes to indicate movements like turning left and right, looking up and down—and capturing facial expressions to generate a dynamic avatar. The avatar, crafted in less than a minute, accurately reflects the user's facial movements, greatly enhancing interactions in video conferencing. **Personal Voice** [2] is another application that creates a synthetic voice that resembles the user's own. To personalize this voice, users are asked to read various texts aloud to generate voice samples. These samples train the system to replicate the user's speech, enabling it to convert text messages into speech that sounds like the user. Moreover, the incorporation of Large Language Models (LLMs), known for their adaptability across diverse linguistic tasks and ability to mimic various speech patterns, introduces a layer of vulnerability. Malicious entities could misuse these features by collecting transcripts, videos, and conversations to analyze a person's speech habits, vocabulary, context preferences, and cultural references. The integration of personalized voice synthesis with the context-sensitivity of LLMs dramatically escalates the potential for identity theft and fraud, posing security and privacy concerns.

With the rising popularity and interest in virtual personal assistants, individuals are becoming more vulnerable to cognitive attacks within hybrid environments. One potential approach to addressing this issue is to develop a multi-modal, discriminative model that welds capabilities in vision, auditory, and language and trains the model from the egocentric perspective, specifically for MR use cases. To improve the robustness of the model, the 'Committee of Experts' approach may utilize diverse sensors, such as audio, video, infrared (IR), ultrasound, thermal, moisture, and vibration sensors, to meticulously verify the authenticity of perceived objects. By integrating data from these diverse sensors, this approach substantially betters the interpretation of the surroundings and markedly lowers the risk of attacks. For example, a 3D model created by Generative AI might duplicate visual aspects accurately, but it lacks the ability to replicate a heat signature, which the proposed model can detect. Furthermore, elevating user consciousness regarding cognitive threats through consistent training to identify various AI model attacks, along with stressing the necessity of regular system updates, serves as an additional layer of defense.

## REFERENCES

[1] Michael Abrash. 2021. Creating the future: Augmented reality, the next human-machine interface. In *2021 IEEE International Electron Devices Meeting (IEDM)*. IEEE, 1–11. https://doi.org/10.1109/IEDM19574.2021.9720526

[2] Apple. 2024. Create a Personal Voice on your iPhone, iPad, or Mac. https://support.apple.com/en-us/104993. Accessed: 2024-02-27.

[3] Apple. 2024. Set up your Persona (beta) on Apple Vision Pro. https://support.apple.com/en-us/HT214002. Accessed: 2024-02-27.

[4] Minghua Liu, Chao Xu, Haian Jin, Linghao Chen, Mukund Varma T, Zexiang Xu, and Hao Su. 2024. One-2-3-45: Any single image to 3d mesh in 45 seconds without per-shape optimization. *Advances in Neural Information Processing Systems* 36 (2024).

[5] Shwetha Rajaram, Franziska Roesner, and Michael Nebeling. 2023. Reframe: An Augmented Reality Storyboarding Tool for Character-Driven Analysis of Security & Privacy Concerns. In *Proceedings of the 36th Annual ACM Symposium on User Interface Software and Technology*. 1–15.

[6] Wen-Jie Tseng, Elise Bonnail, Mark Mcgill, Mohamed Khamis, Eric Lecolinet, Samuel Huron, and Jan Gugenheimer. 2022. The dark side of perceptual manipulations in virtual reality. In *Proceedings of the 2022 CHI Conference on Human Factors in Computing Systems*. 1–15.

[7] Ziqiang Zhang, Long Zhou, Chengyi Wang, Sanyuan Chen, Yu Wu, Shujie Liu, Zhuo Chen, Yanqing Liu, Huaming Wang, Jinyu Li, et al. 2023. Speak foreign languages with your own voice: Cross-lingual neural codec language modeling. *arXiv preprint arXiv:2303.03926* (2023).